# Unveiling Solvent Effects on Femtosecond Laser-Irradiated Au/Fe₃O₄ Colloidal Nanoparticles: The Acetone Effect

Stéphane Mottin[a,b], Żaneta Świątkowska-Warkocka[c], Marta Wolny-Marszałek[c], and Tatiana E. Itina[a*]

[a]*Laboratoire Hubert. Curien, UMR 5516, CNRS, UJM/Univ. Lyon, 18 rue du Prof. Benoît Lauras, Saint-Etienne, 42000, France*

[b]*Laboratoire CREATIS, INSA-Lyon, Univ. Lyon1, CNRS UMR5220, Bâtiment Léonard de Vinci, 21 Av. Jean Capelle, 69100 Villeurbanne, France*

[c]*Institute of Nuclear Physics, Polish Academy of Sciences, ul. Radzikowskiego 152, Krakow, PL-31342, Poland*

**Abstract**

The interplay between laser parameters and liquid environments dictates the outcome of femtosecond laser-induced nanoparticle modification. We present a study of gold and iron oxide nanoparticles in water and a water-acetone mixture, irradiated with femtosecond lasers at 808 nm and 404 nm. While aggregation was observed in pure water at both wavelengths, the results revealed a strong stability and a rather unexpected wavelength-dependency in the acetone-water mixture. In this case, 808 nm irradiation produced some decrease in nanoparticle sizes, while 404 nm led some nanoparticle growth. As a result, the acetone effect is found to be twofold: (i) on one hand, it helps to prevent aggregation; (ii) on the other hand, it acts as a reactive medium allowing to tune the nanoparticle size and composition simply by changing laser wavelength. So, this work emphasizes that solvent physical properties as well as laser-induced chemical processes in the solvent are not merely secondary effects but can dominate the final morphological outcome, providing a predictive framework for nanoparticle synthesis.

# 1. Introduction

The ability to precisely engineer nanomaterials is a cornerstone of innovation in fields ranging from medicine to catalysis [1,2,3,4,5,6,7]. Key challenges and future directions in engineering nanomaterials for medicine and catalysis include achieving precise control over composition and architecture while maintaining biocompatibility, stability, and scalability, as well as understanding multiscale structure–function relationships under realistic physiological or operando conditions. Addressing these issues will rely on advanced in situ diagnostics, data-driven design, and external-field modulation strategies that couple transport, reactivity, and biological response. Laser ablation in liquids [8,9,10,11,12], and more recently femtosecond laser irradiation in liquids [13,14,15,16] has emerged as a versatile tool for such manipulation, capable of inducing nanoparticle (NP) reshaping, sintering, and even fragmentation [17,18,19,20,21,22,23,24,25,26,27,28]. However, predicting the outcome of femtosecond laser-NP interactions in liquids remains challenging due to the complex interplay between laser parameters, NP properties, and the liquid environment [29,30,31,32,33,34]. The stability of nanoparticles in solution is another critical factor influencing their practical applications across various fields, including biomedical and environmental sciences.

The generation of submicro or nano-alloys of immiscible metals is still a challenge using conventional methods. However, because these new materials are currently attracting much attention, alternative methods are needed. In this article, we present a powerful strategy for the generation of a new metastable nanocomposites and alloys of immiscible metals. Recently, bimetallic magneto-plasmonic nanoparticles (NPs) have found numerous applications, in particular as promising contrast agents suitable for bio-imaging and for microand nano-theranostics specially in the domain of oncology [35, 36,37,38] and also for many purposes in the large domain of neurophotonics and brain control in mammals and in birds [39,40,41,42,43]. In fact, these hybrid nanostructures combine magnetic components (such as iron oxide, cobalt or nickel) with plasmonic metals (like gold or silver), enabling manipulation by light and magnetic fields simultaneously. Their bifunctionality allows enhanced imaging modalities including magnetic resonance imaging (MRI) [44,45], magnetic theranostics [46,47] and optical biosensing [48], while also enabling photothermal therapy [49,50] and targeted drug delivery [51,52]. The magnetic core/gold shell structures have demonstrated improved transmigration across biological barriers, and their tunable plasmonic properties in the near-infrared region make them efficient for photothermal cancer treatments [53]. Their multifunctionality supports applications including sensitive molecular imaging, magnetic hyperthermia, and dual-modal magnetic/optical imaging, making them highly attractive for oncology theranostics [54,55,56,57,58]. Specifically, in the case of brain cancer, treatment options include many choices (surgery, chemotherapy, radiation and also targeted therapy i.e. VEGF inhibitors) [59,60,61], but still remain challenging and not effective enough because of their locations making it hard for surgery, but also because of many possible complications, collateral damage, etc. Thus, we believe that these new magneto-plasmonic NPs will pave the way toward not only the new end efficient diagnostics and imaging, but also to the novel and more precise treatment options based on their unique properties.

Aggregation, sintering, and coalescence are extremely important in nanoparticle synthesis. These terms, however, describe distinct physical processes that describe the behavior of particles under different conditions. While aggregation refers to the process where individual particles or molecules come together to form larger clusters without significant change in their internal structure, sintering involves the fusion of particles at high temperatures, where atomic diffusion leads to the reduction of porosity and the formation of a solid mass. Coalescence, on the other hand, describes the merging of two or more particles into a single larger entity, typically driven by surface energy minimization. Each of these processes plays a critical role in material properties and behavior under various environmental conditions. For example, aggregation can be reversable and is often observed in colloids at moderate temperatures. Laser-induced nanoparticle sintering is a non-reversable process, which is crucial in direct three-dimensional (3D) laser manufacturing, or 3D printing [62]. These processes have recently attracted a great attention due to their capacity of creating nanoobjects for optics, photonics and for medicine. Finally, nanoparticle coalescence often takes place upon strong heating and melting so that nanoparticles can be considered as nanodroplets. Coalescence can be helpful in creation nanoalloys or core-shell nanoparticles. A better understanding of the physical and chemical processes involved in such processes has a potential of promoting many applications of this laser-based technology [63,64].

A critical, often underexplored factor in laser-assisted metallic composite and alloy nanoparticle behavior is the role of the solvent itself. Beyond acting as a heat sink, the solvent can undergo laser-induced ionization and decomposition, generating reactive species and transient states that drastically alter NP stability and reactivity [65,66,67,68]. In particular, oxidation can become important considerably modifying magneto-optical properties of nanoparticles [69,70,71,72]. Additionally, several recent theoretical studies focused on laser interactions in water have shown that intense short and ultra-short pulses can drive its multiphoton ionization and sometimes its dissociation [73,74,75,76], leading to the formation of micro- or nanoplasmas, creating conditions of optical breakdown at high pulse energies [77], but also often for micro- and nanobubble formation [78,79,80,81,82], solvation of electrons [83], molecular dissociation and a cascade of radical species (•OH, •H, $e_{aq}^-$) at higher laser intensities [84], etc. At lower laser intensities, however, ultrafast laser pulses can also also result into chemical modifications of nanoparticles that can be strongly wavelength and solvent-dependent. If multi-pulse irradiation is used, furthermore, an ensemble of nanoparticle can experience a set of altering modifications including melting, sintering and aggregation [67].

In organic solvents like acetone, laser irradiation can induce decomposition pathways that creates active carbon capable to form solid carbonaceous deposit on nanoparticles [85,86,87]. Some nanoparticles can even serve as catalysis and promote this reaction. In all these processes, laser wavelength is also a key parameter, as it determines the efficiency of both direct NP excitation and solvent ionization. For example, 400 nm irradiation is expected to be more efficient at exciting plasmonic NPs and driving plasmon-mediated photochemical reactions.

This study examines the role of solvents in femtosecond laser-irradiation of colloidal nanoparticles. In particular, we investigate the behavior of gold (Au) and magnetite ($Fe_3O_4$) nanoparticles suspended in water versus a water-acetone mixture when subjected to irradiation at wavelengths of 404 nm and 808 nm. The introduction of acetone alters the solvent characteristics, and this environment significantly impacts the aggregation behavior of the nanoparticles. The article is organized as following. Experimental details are presented in Section 2, experimental results are given in Section 3, three molecular dynamics models and computational results are presented in Section 4. Section 5 gives a short discussion of experimental results. Conclusions are summarized in Section 6.

## 2. Experimental Details

2.1 Materials

**Nanoparticles**. Both gold (Au NP's) and iron oxide nanoparticles ($Fe_3O_4$ NPs) were formed by femtosecond PLAL in LP3 Laboratory, Marseille, and were stable enough without any additional ligands. Gold nanoparticles nominal diameter was 25 nm. The solutions had a concentration of 0.8 mg/mL, as determined by the supplier using spectrophotometry. The as-received Au NPs were stabilized in citrate buffer solution. Aqueous suspension of iron (III) oxide nanoparticles (magnetite, $Fe_3O_4$) had nominal diameter of 30-50 nm. The concentration of the solution was < 0.05 mg/mL, as stated by the supplier. The $Fe_3O_4$ NPs were stabilized in water and were magnetically separated before mixing with the Au NPs solution.

**Solvents.** Ultrapure water (resistivity > 18.2 MΩ·cm) obtained from a Millipore Milli-Q water purification system was used for all aqueous solutions. Acetone (ACS grade, purity ≥ 99.5%). A 50:50 (v/v) water-acetone mixture was prepared by mixing equal volumes of ultrapure water and acetone.

2.2 Methods

**Nanoparticle Mixture Preparation**. Prior to laser irradiation, the Au and $Fe_3O_4$ NP stock solutions were mixed in a 1:1 volume ratio (125 µL Au NPs + 125 µL $Fe_3O_4$ NPs). This mixture was then diluted either with 250 µL of ultrapure water or 250 µL of the 50:50 water-acetone mixture, resulting in a 4-fold dilution of the original stock solutions.

**Femtosecond Laser Irradiation.** The nanoparticle suspensions were irradiated from the side by using an amplified Ti:Sapphire laser system (Thales Laser, Concerto) operating at a central wavelength of 808 nm with a pulse duration of 130 fs and a repetition rate of 1 kHz. The laser beam was frequency-doubled using a beta-barium borate (BBO) crystal to generate the second harmonic at 404 nm. The average laser power was measured before the irradiation cell using a power meter (Thorlabs PM100D). The laser power at 808 nm was 160 mW, and at 404 nm was 60 mW.

**Experimental Setup.** The nanoparticle suspensions were placed in a 1 cm path length quartz cuvette (Hellma Analytics). The cuvette was positioned perpendicular to the laser beam. The laser beam was focused onto the cuvette using a quartz lens with a focal length of 100 mm for 808 nm irradiation and 50 mm for 404 nm irradiation. The focal spot was located approximately 4 mm behind the front face of the cuvette. The laser was operated in the TEM00 mode and the laser spot size was approximately 4x4 $mm^2$.

**Irradiation Procedure.** Each sample was irradiated by femtosecond laser pulses for 1 hour under continuous stirring to ensure uniform exposure.

**UV-Vis-NIR Spectroscopy**: The optical absorption spectra of the nanoparticle suspensions were recorded before and after laser irradiation using a Cary 7000 spectrophotometer (Agilent). Spectra were acquired over a wavelength range of 300-800 nm with a spectral resolution of 1 nm. The baseline was corrected using the appropriate solvent as a reference.

**Scanning Electron Microscopy (SEM):** The morphology of the nanoparticles was investigated using a TESCAN Vega-3 scanning electron microscope. Following laser irradiation, a small aliquot of each sample was drop-cast onto a silicon wafer and allowed to air dry. The samples were then sputter-coated with a thin layer of gold to enhance conductivity. SEM images were acquired at an accelerating voltage of 30 kV.

3. Experimental Results

To investigate the interplay between laser parameters, nanoparticle properties, and the liquid environment, our study strategically employs two distinct femtosecond laser wavelengths: 808 nm and 404 nm. This selection is crucial for differentiating between direct nanoparticle heating effects and solvent-mediated processes, particularly in relation to the gold nanoparticle (Au NP) surface plasmon resonance (SPR) and the multiphoton absorption properties of the solvents. The gold nanoparticles used in this study exhibit a surface plasmon resonance (SPR) at approximately 520 nm. This SPR correspond to the collective oscillation of conduction electrons in response to incident light, leading to enhanced light absorption and localized heating.

Thus, the 808 nm wavelength lies significantly off-resonance of the Au NPs. Consequently, direct photon absorption and subsequent photothermal heating of the nanoparticles are expected to be comparatively less efficient. Femtosecond pulses at this wavelength, however, can still deposit substantial energy into the colloids due to non-linear effects, such as "white continuum" generation. Herein, using two wavelengths, 404 and 808 nm, allows us to isolate and study solvent-mediated effects such as photochemical reactions. In fact, direct nanoparticle absorption and plasmon-mediated effects are negligible at 808 nm, and the lower photon energy makes it insufficient to efficiently drive the photochemical decomposition in the solvents. On the contrary, because 404 nm is much closer to the SPR of the Au NPs, direct energy absorption in nanoparticles (particularly, gold) is enhanced leading to highly efficient photothermal heating of the nanoparticles. The higher photon energy associated with 404 nm also plays a critical role in promoting photochemical reactions within the solvent environment. Specifically, it is efficient at initiating the photodecomposition of acetone into carbonaceous byproducts. Thus, by employing 404 nm, we could create conditions where both efficient direct nanoparticle heating and solvent chemical reactivity (which can lead to sintering via carbon deposition) are prominent.

Below we systematically compare these two distinct wavelengths across pure water and water-acetone mixtures.

3. 1 Scanning Electron Microscopy (SEM)

The average particle size in samples irradiated at 404 nm in the water-acetone mixture increased by several orders of magnitude compared to the unirradiated samples and the samples irradiated in ultrapure water. Representative SEM images of the nanoparticle suspensions after laser irradiation are shown in Figures 1-4.

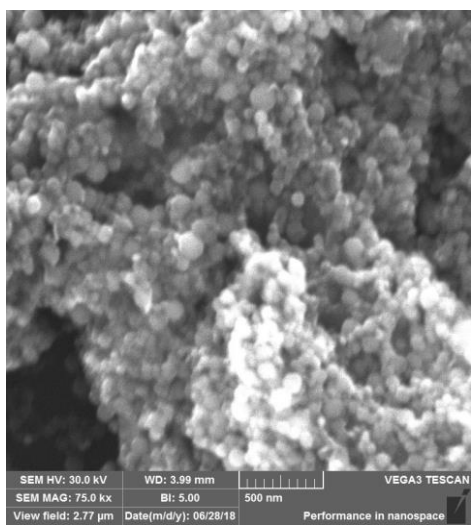

Figure 1. Image of nanoparticles after 60 min of irradiation at 404 nm in water showing prevailing aggregation of particles

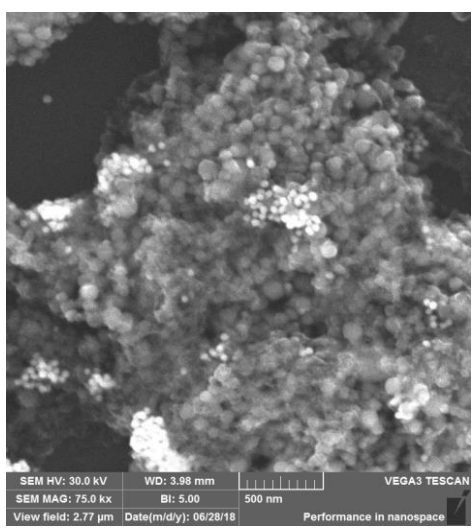

Figure 2. Image of nanoparticles after 60min of irradiation at 808 nm in water showing aggregation of particles

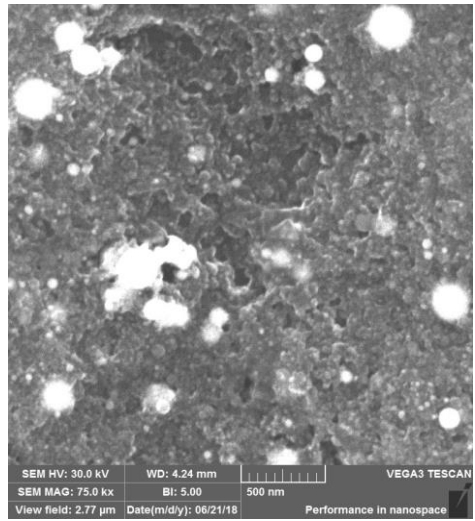

Figure 3. Image of nanoparticles after 60 min of irradiation at 404nm in 50% water-50% acetone showing sintering and coalescence of nanoparticles.

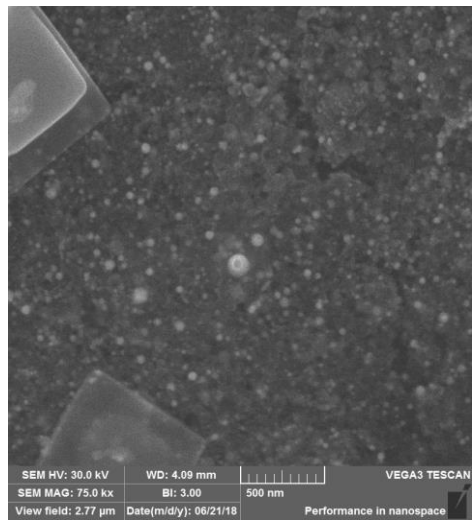

Figure 4. Image of nanoparticles after 60min of irradiation at 808 nm in 50% water-50% acetone showing fragmented and smaller particles (there are two artefacts with cubic crystallization due the solvation by acetone).

The solvent environment probably played a critical role in determining the outcome of laser irradiation. In more details:

**Irradiation in Ultrapure Water.** SEM images of samples irradiated at 808 nm in ultrapure water showed a relatively uniform distribution of nanoparticles with many aggregates formed from the particles with sizes slightly smaller than the initial ones compared to the unirradiated samples. Surprisingly, the ultrafast laser irradiation at 404 nm in ultrapure water also revealed nanoparticle aggregation rather than fragmentation.

**Irradiation in 50:50 Water-Acetone Mixture.** SEM images of samples irradiated at 808 nm in the water-acetone mixture revealed a noticeable colloidal stability. The nanoparticles remained relatively well-dispersed, with no significant aggregation. Laser irradiation at 404 nm in the water-acetone mixture also led to the absence of strong aggregation too, but revealed some changes in nanoparticle's morphology suggesting some effects of melting [67].

3.2 UV-Vis-NIR Spectroscopy

The results of UV-Vis-NIR spectroscopy are generally consistent with SEM imaging. Figures 5-7 present the UV-Vis-NIR absorption spectra of Au and $Fe_3O_4$ nanoparticle suspensions before and after femtosecond laser irradiation at 808 nm and 404 nm in both ultrapure water and the 50:50 water-acetone mixture.

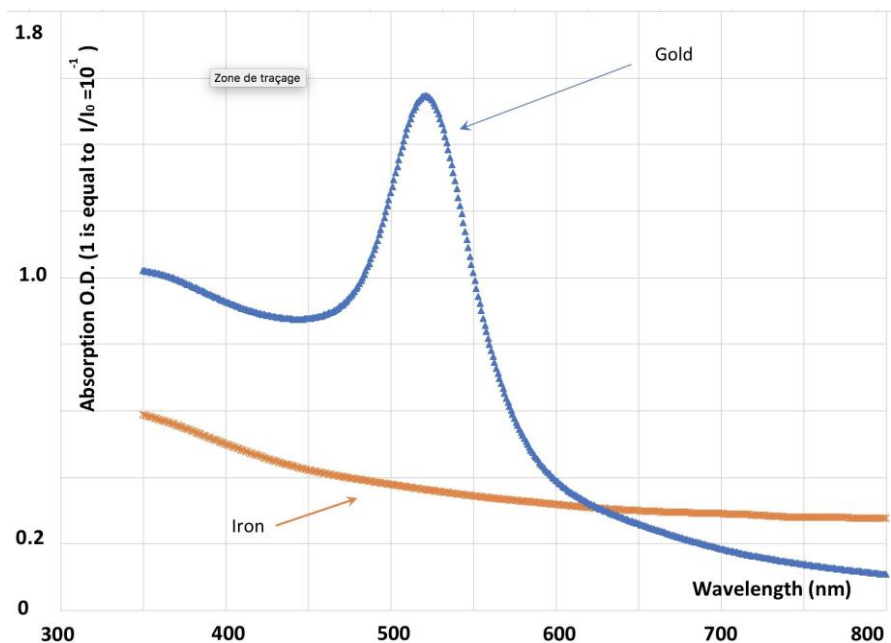

Figure 5. UV-Vis-NIR spectra of the unirradiated Au and $Fe_3O_4$ nanoparticle solution in water, showing the Au SPR peak at ~520 nm and broad $Fe_3O_4$ absorption.

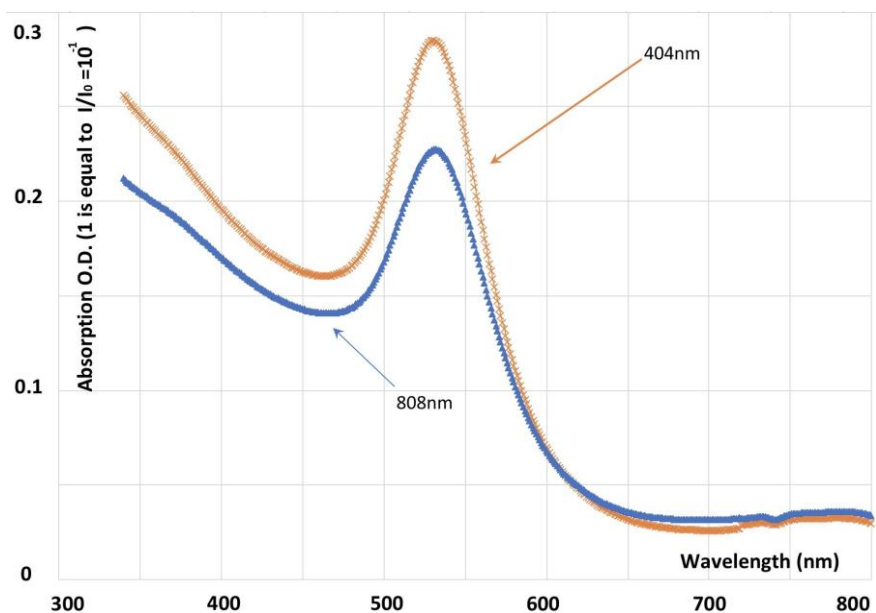

Figure 6. UV-Vis-NIR Spectra of Au and $Fe_3O_4$ nanoparticle solutions in water after laser irradiation (404nm and 808nm).

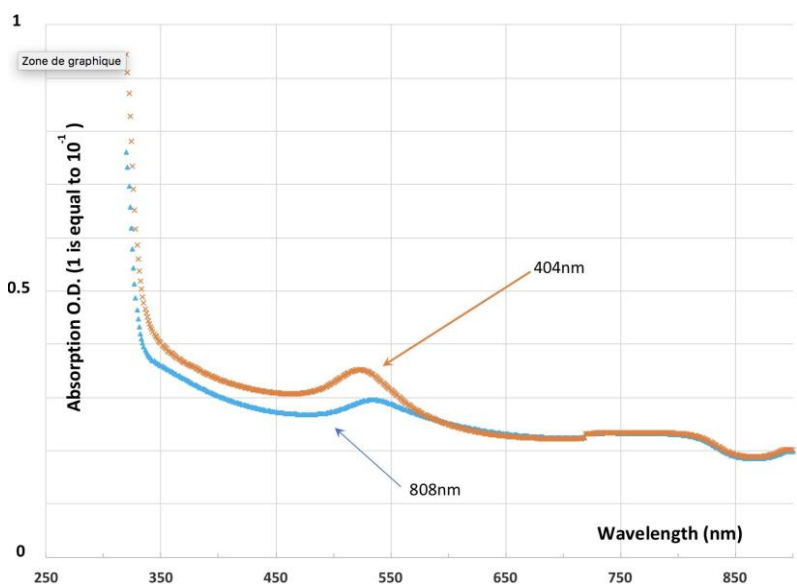

Figure 7. UV-Vis-NIR Spectra of Au and $Fe_3O_4$ nanoparticle solution in 50% water-50% acetone after 60min laser irradiation (404nm and 808nm).

**Unirradiated Samples.** Prior to laser exposure, all samples exhibited a characteristic absorption peak at approximately 520 nm, corresponding to the surface plasmon resonance

(SPR) of the 25 nm gold nanoparticles. The iron oxide nanoparticles contributed to a broad absorption feature in the UV region, with a decreasing absorption towards the NIR region.

**Irradiation in Ultrapure Water.** After irradiation at 808 nm in ultrapure water, the SPR peak at 520 nm decreased in intensity by approximately 15% (± 3%), indicating a reduction in the concentration of the original Au NPs. A very slight blue shift (≈ 5 nm) of the SPR peak was also observed indicating some decrease in the sizes of unit Au NPs that could, however, aggregate with iron oxide nanoparticles. As for the laser irradiation at 404 nm in ultrapure water, it resulted in a more pronounced decrease in the SPR peak intensity (approximately 30% ± 5%) and a blue shift (≈ 10 nm).

**Irradiation in 50:50 Water-Acetone Mixture.** After irradiation at 808 nm in the water-acetone mixture, the SPR peak showed a slight decrease in intensity (approximately 10% ± 2%) and a minimal shift in peak position (≤ 2 nm). In contrast to the results obtained in ultrapure water, irradiation at 404 nm in the water-acetone mixture led to a red shift of the SPR peak, shifting from 520 nm to approximately 545 nm. The peak intensity also decreased by approximately 20% (± 4%). The red shift observed after irradiation at 404 nm in the water-acetone mixture suggests a slight increase in the average size of the nanoparticles without strong aggregation.

## 4. Modeling and Computational Results

We start by calculation extinction, absorption and scattering efficiencies for gold and Fe2O3 nanoparticles. Here, classical Mie theory was used. The obtained results are shown in Figure 8 (a-c) revealing well-pronounced plasmonic peak for gold nanoparticles. We note that when a liquid solvent is present, this peak is slightly shifted to the right as compared to vacuum or air value and depends on nanoparticle size (517 nm for 20 nm diameter and 527 nm for 50 nm diameter in water at room temperature). These simulation results raters well correlate with the above presented experimental observations. As expected, the absorption of gold nanoparticles is much higher at 404 nm because it is closer to the plasmon resonance value, while 808 nm is out of the resonance. Iron oxide nanoparticle also better absorb at 404 nm, even though they do not have any plasmon resonance.

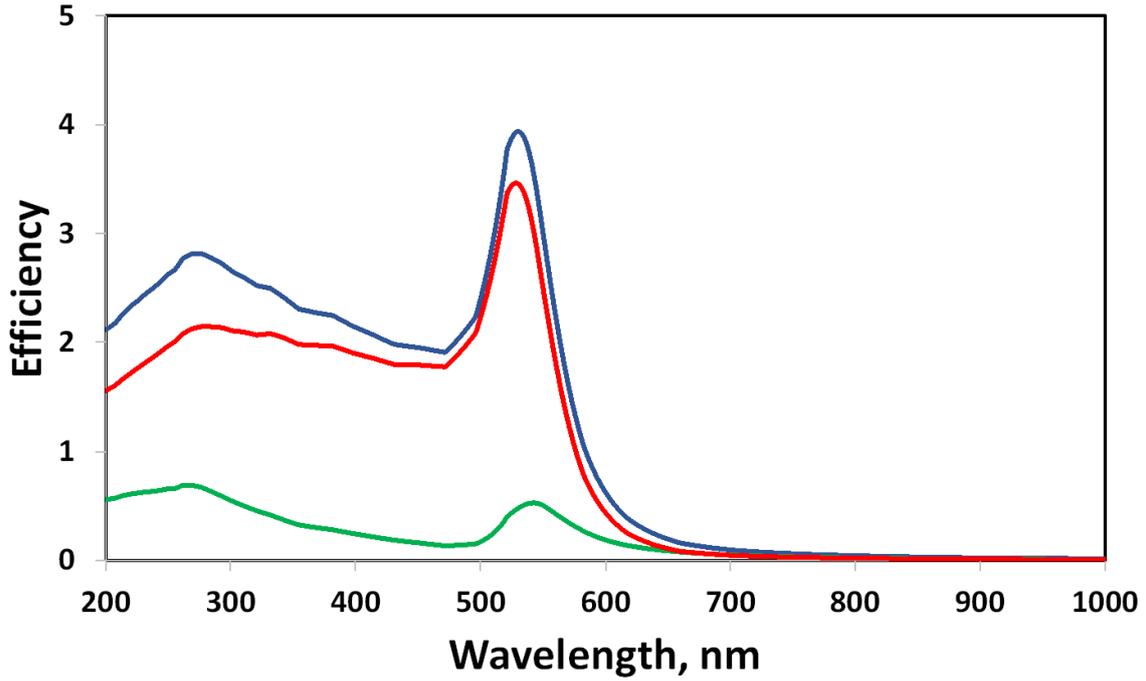

Figure 8. Calculated absorption, scattering and extinction efficiency Q=σ/(π$R_{np}^2$) of gold nanoparticles with d=50 nm in water at room temperature. Here, σ is the corresponding cross-section, $R_{np}$ =d/2 is nanoparticle radius; blue line corresponds to extinction efficiency; red line to absorption efficiency; and green line shows 3nanoparticle scattering efficiency.

The temporal evolution of nanoparticle and solvent temperatures following a single femtosecond laser pulse was meticulously modeled using a three-temperature model (3TM), an extension of the classical two-temperature model (TTM) that describes laser energy absorption by the electron subsystem and its subsequent transfer to the phonon (lattice) subsystem via electron-phonon coupling [23,27, 33]. The 3TM introduces a third temperature equation for the surrounding solvent, thereby accurately depicting the intricate heat dynamics encompassing both internal nanoparticle heating and energy transfer to the external medium. The system of coupled differential equations governing the electron $T_e$, lattice ($T_l$), and medium ($T_m$) temperatures is given by equations (1-3) as follows

$$\frac{dT_e}{dt} = \frac{1}{C_e(T_e)}\left(-G(T_e)(T_e - T_l) + S(t)\right) \quad (1)$$

$$\frac{dT_l}{dt} = \frac{1}{C_l(T_l)}\left(G(T_e)(T_e - T_l) - \frac{3h(T_l - T_m)}{R_{np}}\right) \quad (2)$$

$$\frac{dT_m}{dt} = \frac{1}{C_m}\left(\frac{3h(T_l-T_m)}{R_{np}}\right) \quad (3)$$

Here, $S(t) = \frac{\sigma_{abs}I(t)}{V_{np}}$ represents the volumetric heat source from the laser, where $I(t)$ is the laser intensity and $V_{np}$ is the nanoparticle volume. $G(T_e)$ is the electron-phonon coupling constant, and $C_e$, $C_l$, and $C_m$ are the volumetric heat capacities of the electron, lattice, and medium, respectively. The term $h$ denotes the interfacial heat transfer coefficient, which inherently accounts for Kapitza resistance, the thermal boundary resistance that impedes heat flow across the solid-liquid interface at the nanoscale. By using this model, calculations were performed for gold nanoparticle with a radius $R_{np}$ of 25 nm. Material parameters for gold included a density of 9300 kg/m³, a molar mass of 0.19696657 kg/mol, and a volumetric electronic heat capacity coefficient $Ce$ derived from a standard molar value of 6.74*10⁻⁵ J/(mol K²). The electron-phonon coupling parameter $G$ was set to here 1*10¹⁷ W/m³K. All subsystems were initialized at 300 K. The nanoparticle was irradiated by a 404 nm femtosecond laser pulse (130 fs duration, 300 J/m² fluence with a repetition rate of 1 kHz). A Mie absorption efficiency ($Q_{abs}$) of 1.885 was used, yielding an absorption cross-section ($\sigma_{abs}$) of approximately 3.7 10⁻¹⁵ m². The interfacial heat transfer coefficient, $h$, from the gold nanoparticle to the solvent was approximated as ~105.106 W/(m²K), a value typically reported for gold-water interfaces. It is important to note that while this $h$ value encapsulates the Kapitza resistance for the gold-water interface, its magnitude for a gold-acetone interface could differ due to varying interfacial phonon coupling and solvent properties. Volumetric heat capacities ($C_m$) for the surrounding media were 4.184*10⁶ J/(m³K) for water and 1.696*10⁶ J/(m³K) for acetone (derived from its density of 789 kg/m³ and specific heat of 2150 J/(kg K).

The performed calculations demonstrated that following the femtosecond laser pulse, the electron temperature ($T_e$) within the gold nanoparticle rapidly surged to an extreme peak of approximately 31,804 K, subsequently transferring energy to the lattice. The electron-phonon energy exchange led to the nanoparticle's lattice temperature $(T_l)$ peaking at about 1,475 K within approximately 20 ps, significantly exceeding gold's nanoparticle size-dependent melting point (here, near 1337 K). Despite these drastic internal nanoparticle temperature increases, the observed temperature change in the adjacent solvent ($T_m$) remained relatively modest. When our simulations tracked temperatures over the initial 100 ps following the pulse, the results showed that the temperature of the surrounding water experienced a maximum rise of 0.005 K, while the acetone medium demonstrated a maximum rise of 0.012 K. To observe a more advanced local thermal equilibration between the nanoparticle and its immediate solvent environment, calculations were then extended to 1 nanosecond. At this longer timescale, the temperature of the surrounding water showed a total rise of 0.046 K, whereas the acetone medium exhibited a total rise of 0.114 K. This trend, where the temperature rise in acetone was approximately 2.5 times greater than in water, is directly attributable to acetone's substantially lower volumetric heat capacity 1.696 *$10^6$ J/(m$^3$K) compared to water (4.184*$10^6$ J/(m$^3$K)). Regarding the laser's 1 kHz repetition rate, the time between pulses is 1 ms. While the nanoparticle-solvent local equilibration occurs within nanoseconds (for the nanoparticle) to microseconds. For the immediate solvent shell, with a characteristic thermal diffusion time $\tau_d \approx R_{eff}^2/\alpha_{sol}$, where $R_{eff}$ is the heated zone radius and $\alpha_{sol}$ is the thermal diffusivity. This microsecond cooling time is significantly shorter than the 1 ms inter-pulse duration. For instance, for water, the thermal diffusivity is approximately 1.4 * $10^{-7}$ m$^2$, implying a heated region would cool in about 5 µs. This suggests that the heated volume around the nanoparticle largely dissipates its energy into the bulk solution before the arrival of the next pulse, preventing substantial local peak temperature accumulation from pulse to pulse. However, over extended irradiation periods, the cumulative effect of many pulses can still lead to a gradual, macroscopic increase in the average temperature of the entire solution. Linearly extrapolating from the estimated local temperature rises, we get approximately 256 laser pulses would be required for the acetone medium, initially at 300 K, to reach its boiling point of 329.15 K (56 °C). Under the same conditions, the corresponding local water temperature would reach approximately 311.76 K (only 38.61 °C), while our water-acetone mixture would reach around 314.85 K (41.7 °C). The marked difference in temperature accumulation is directly attributable to acetone's substantially lower volumetric heat capacity compared to water. For a 50%x50% water-acetone mixture, the local temperature rise per pulse is estimated to be approximately 0.0655 K, reflecting its intermediate volumetric heat capacity between pure water and pure acetone. Consequently, it would require around 445 such laser pulses for the mixture to reach the boiling point of pure acetone at which point complex phase change dynamics would initiate. Importantly, when acetone reaches its boiling point, a critical concentration difference emerges since liquid rapidly converts to a highly concentrated vapor phase, with minimal dissolved gases. Thus, in a 50/50 water-acetone mixture at 329.15 K, preferential evaporation of the more volatile acetone takes place. The density difference between a liquid phase and its saturated vapor is commonly several orders of magnitude, illustrating a phase transition.

To further investigate the observed nanoparticle aggregation behavior, particularly its dependence on local solvent concentration, we performed a set of molecular dynamics (MD) simulations using LAMMPS (Figure 9). The simulations were conducted in two dimensions using Lennard-Jones (LJ) reduced units, with periodic boundary conditions. Four types of particles were modeled to represent the colloidal system: a single central Brownian particle (Type 2, mass 50 LJ units), two species of other colloidal particles (Type 3 with mass 45 LJ units, and Type 4 with mass 15 LJ units), and numerous background particles (Type 1, mass 1 LJ unit) representing the solvent molecules. In these simulations, a soft potential was defined within a cutoff distance $r_{cut}$. The interaction between particles of types i and j was given by the following equations: $V_{ij}(r) = e_{ij} (1 - r/l_{ij})^2$, $r < r_{cut}$, and $\epsilon_{ij}$ is the interaction strength between particles of types i and j, (e.g., $\epsilon_{11} = 10.0$, $l_{11} = 1.0$ for solvent-solvent interactions, see Supplementary Materials for full parameter sets). A common interaction cutoff of 6.0 LJ length units was applied. The initial kinetic energy of the system corresponded to a reduced temperature of 2.0 in LJ units (velocity all create 2.0). To simulate the thermal conditions corresponding to our laser-heated solvent environments, the system temperature was maintained using an NVT thermostat. This LJ temperature of 29.60 effectively corresponds to an elevated real-world temperature of approximately 314.85 K (41.7 °C), representing the nanoparticle-heated environment. Each simulation runs for 500,000 timesteps, with each timestep corresponding to 0.001 ps.

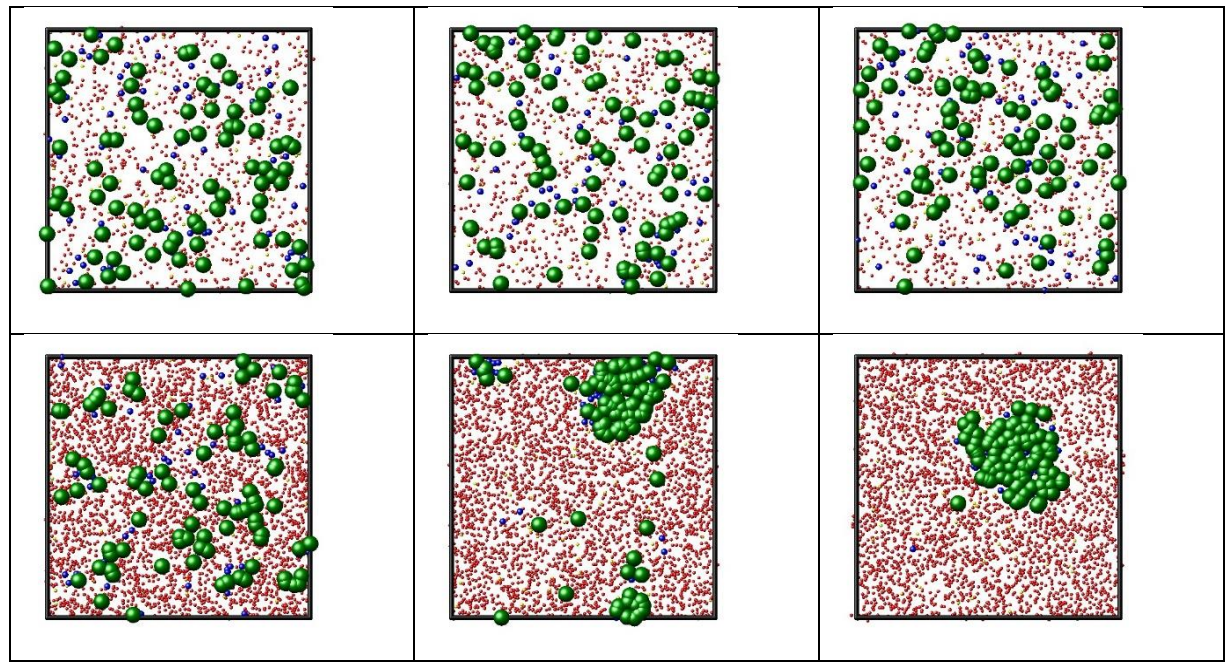

Figure 9. The obtained calculation results at the beginning of the simulation, at the middle and at the end for two different solvent concentrations (upper road is for the lower one and the second on is for higher concentration at three time delays).

The impact of local solvent concentration was directly explored by varying the number of these Type 1 solvent particles. In the first simulation, 800 particles representing a less dense solvent environment, while 3000 were used to represent a denser solvent in the second simulation (an approximately 3.75-fold increase in solvent particle count for the same box volume). As one may see in Figure 9, the performed MD simulations consistently demonstrated that nanoparticle aggregation is significantly more pronounced in a solution with a higher local solvent concentration, fostering a denser medium for nanoparticle interaction and promoting aggregation. This finding aligns with the obtained experimental results, where in acetone-water mixture aggregation was less pronounced than in a denser water solvent.

## 5. Discussion

The impact of the solvent on the behavior of nanoparticles under femtosecond laser irradiation is critical in understanding both their stability and response to laser energy. In this study, the colloidal nanoparticles were irradiated in an acetone-water mixture (50%:50%) and pure water under similar conditions. Despite the identical laser intensities and wavelengths used (404 nm and 808 nm), the nanoparticles in the acetone-water mixture exhibited no noticeable aggregation, whereas nanoparticle aggregation was observed in pure water. This contrast can be attributed to several key factors related to solvent interactions, the nature of the nanoparticles, and the laser-induced effects.

Acetone, as a polar solvent, has a strong affinity for disrupting intermolecular forces between particles and altering the solvation shell of nanoparticles. In a water-acetone mixture, the presence of acetone reduces the hydrogen bonding network typically seen in pure water. Water's highly structured hydrogen bond network plays a critical role in nanoparticle aggregation, as it stabilizes the electrostatic and van der Waals interactions between particles. When acetone is introduced, its lower polarity and ability to disrupt hydrogen bonds weaken these interactions, thereby preventing the aggregation that would otherwise occur in water alone. This solvation effect can be modeled using the following relationship for the interaction potential between nanoparticles in solution: $V = V_{el} + V_{VdW} + V_{solv}$, where $V_{el}$ and $V_{VdW}$ correspond to the colloidal forces between particles (commonly, Von der Waals forces), and $V_{solv}$ accounts for the effects of the solvent. In the case of acetone-water mixtures, the reduced solvation energy weakens the overall interaction potential, which favors the maintenance of nanoparticle dispersion. The solvation potential $V_{solv}(r)$ can be expressed using models such as the Polarizable Continuum Model (PCM) or dielectric continuum models, which account for the solvent's dielectric constant and its effect on nanoparticle surface charge. For simplicity, let us focus solvation energy change $\Delta G_{solv}$ which is generally refers to the energy associated with the interaction between the nanoparticle surface and the solvent molecules, which influences the stability of a colloidal suspension. It can be estimated can be estimated using the Born equation for a spherical nanoparticle as follows $\Delta G_{solv}=-A/\varepsilon$, where $A$ is the effective surface area interacting with the solvent (for spherical particles totally covered by the solvent, $A = 4\pi r^2$) ; and $\varepsilon$ is the dielectric constant of the solvent determining the strength of the solvation energy ($\varepsilon$ =78.5 for pure water and 20.7 for acetone). The higher this constant, the stronger the solvent can stabilize charged or polar nanoparticles. For nanoparticles in water ($r$=25 nm), $\Delta G_{solv}$ =-4 $10^{-17}$ J, while in 50%x50% acetone-water mixture, $\varepsilon$=(78.5+20.7)/2= 49.6 giving $\Delta G_{solv}$= -2.52 $10^{-17}$ J. As a result, the solvation energy in the acetone-water mixture is weaker implies that the overall repulsive forces between nanoparticles are reduced.

While the Born equation suggests weaker electrostatic stabilization in acetone-water mixtures due to their lower dielectric constant, the obtained experimental data, however, indicates a reduced aggregation in this case. This discrepancy clearly demonstrates the limitations of purely continuous electrostatic models for ligand-coated nanoparticles in reactive solvents. To explain this rather unexpected result, we note that the differences in viscosity, density, concentration, thermal conductivity, solvent structure between water and the acetone-water mixture can explain the observed differences in aggregation behavior. Additionally, electro-magnetic field generated by the femtosecond laser pulse can induce rapid heating at the nanoparticle surface, leading to a localized increase in temperature and, in some cases, promoting photo-chemical and surface reactions.

Firstly, as one can see from simulation results, an overall decrease of solvent density or in particle local concentration evidently prevents aggregation. Since acetone-water mixture is less dense than water, less aggregates are expected to be formed. Rapid evaporation of acetone may also lead to nano- and microbubble formation around the nanoparticles. In fact, acetone, with a boiling point of 56°C, evaporates faster than water (boiling point of 100°C). This rapid bubble formation can disrupt the aggregation process. The optical absorption of both Au and $Fe_2O_3$ nanoparticles is, however, higher at 404 nm than at 808 nm (Figures 8 for Au NP).

Water molecules are known to form a highly structured hydrogen bonding network. This network can create a "cage" or structured layer around nanoparticles. While this can provide some initial stability, it can also lead to aggregation if these water layers become disrupted or if inter-particle hydrogen bonding (or bridging by water molecules) becomes favorable during collisions, especially when localized heating occurs. The highly structured water can make it harder for particles to escape close contact once they have overcome repulsion. On the other hand, acetone is a polar, aprotic solvent. Its introduction disrupts the strong hydrogen bonding network of water. This disruption means the solvent environment around the nanoparticles is less structured. This reduction in solvent structuring can effectively make the solvent less "sticky" and reduce water-mediated attractive forces that might contribute to aggregation in pure water. It is akin to a steric effect where the solvent itself becomes a less effective medium for particles to "stick" together. Additionally, the interaction of a polar solvent like acetone with the nanoparticle surface can alter the surface charge and/or surface chemistry. Acetone molecules could adsorb onto the nanoparticle surface, modifying the double layer and thus influencing electrostatic repulsion. This could lead to an increase in surface charge or create a different repulsive layer compared to water.

At high laser intensities, furthermore, interactions may be dominated by multiphoton absorption. Electrons can be injected into the solvent, promoting avalanche ionization, ambient heating, and cavitation. These effects are, however, more likely to occur at 404 nm and require tight laser focusing, which was not the case in our experiments.

To better explain the wavelength-dependent effects, it is crucial to consider photo-chemical reactions. For instance, acetone may undergo decomposition under the required laser irradiation, potentially forming carbonaceous byproducts. These species could adsorb onto the nanoparticle surface, creating a protective layer that either physically prevents aggregation (via a steric barrier) or chemically modifies the surface to enhance repulsion. The presence of such a layer can directly counteract aggregation, modifying particle surface tension, surface charge and zeta potential. In this way, a possible explanation was proposed for the acetone effect in stable gold nanoparticle formation in pulsed laser ablation experiments with 25 ps at 532 nm wavelength).[88] In particular, as it was previously suggested, the stabilization mechanism in acetone can be related to the light-induced formation of enolate on the gold surface. In some cases, this enolate can undergo degradation with formation of amorphous carbon, that was confirmed by surface enhanced Raman spectroscopy (SERS). Additionally, a sharp colloidal absorption was observed and a so-called « photo-acetone method » was proposed at the wavelength of ~400 nm for preparation of fine metal particles[89]. It was applied to obtain colloidal silver, gold, copper and platinum, and the main effect was attributed to the formation of highly reactive $(CH_3)_2CO^*$ radicals from $(CH_3)_2COH$. Furthermore, 404 nm light can promote the reduction of dissolved metal ions potentially causing Ostwald ripening or growth[90]. The photon energy at 404 nm (~3.1 eV) is sufficient to drive photochemical reduction reactions. In this case, acetone radicals could act as reducing agents. Conversely, the photon energy at 808 nm (~1.5 eV) is too low to drive this reaction.

The obtained results highlight the delicate balance between laser-induced heating, solvent interactions, and nanoparticle stability. While the laser pulse provides the initial energy input, it is the solvent properties—such as polarity, thermal conductivity, and reactivity—that determine the long-term behavior of nanoparticles, including their aggregation tendencies. Understanding these interactions is crucial for controlling the stability and functionality of colloidal systems, especially in applications where nanoparticle aggregation could compromise performance, such as in numerous plasmonic applications, solar cells and catalysis.

## 5. Conclusions

This study demonstrates that the solvent environment is not a passive bystander but an active participant in femtosecond laser-induced nanoparticle modification. Interestingly, under irradiation at 808 nm in 50:50 water-acetone mixture the nanoparticles remained relatively well-dispersed, with no significant aggregation. Laser irradiation at 404 nm in the water-acetone mixture led to the formation of a bit larger spherical particles due to sintering. In pure water, however, aggregation was observed for both 404 and 808 nm. The addition of acetone to the water solvent significantly alters the outcome of the laser processing, with acetone acting as an active chemical participant in the reaction. In pure water, nanoparticles tended to aggregate and form clumps, regardless of the laser wavelength used. In this article, we introduce the concept of the "acetone effect." When acetone is added, aggregation is almost entirely inhibited, and the mixture provides enhanced stability, effectively keeping the nanoparticles separated.

The contrasting behaviors of nanoparticles in water and in an acetone-water mixture under femtosecond laser irradiation can be attributed to the differing effects of these solvents on nanoparticle interactions. In pure water, higher concentration may enhance aggregation, whereas in the acetone-water mixture, the rapid solvent evaporation at lower temperature and nano- and microbubble formation can drive nanoparticle dispersion by disrupting aggregation processes. These effects, however, do not explain the observed wavelength dependency, that can be attributed mostly to photo-chemical effects induced by the laser. Particularly, such processes take place much more likely at 404 nm, contributing to the observed differences in nanoparticle behavior. At this wavelength, the stabilization mechanism in acetone can be related to the light-induced formation of enolate on the gold surface. Additionally, acetone radicals can be formed to reduce metal ions enhancing nanoparticle growth instead of the aggregations, whereas the lower photon energy at 808 nm is insufficient for such reactions. The obtained results highlight the importance of both solvent properties and laser wavelength in controlling nanoparticle dynamics.

These insights provide a predictive framework for designing laser-based nanofabrication protocols. By selecting the appropriate solvent and laser wavelength, one can better control the synthesis of advanced nanomaterials with tailored architectures, optimizing their performance for applications in catalysis, sensing, 3D printing, solar cells, and medicine.

**Authors Contributions**

TEI, ZSW supervised the project, obtained funds, and performed the analysis of the results. SM conceived the experimental set-up. ZSW and SM performed a series of experiments. TE performed modeling. MWM initiated the discussions. All the authors have participated in the manuscript preparation.

**Acknowledgements**

The authors gratefully acknowledge the financial support provided by the French-Polish collaborative PHC Polonium project (Project No. 47829ZL), operated by CAMPUS FRANCE, and partially from ANR PEPR LUMA SUNRISE. The authors would like to thank their colleagues from LP3 laboratory in Marseille, France (in particular, the group of Dr. Andrei V. Kabashin), for the supply of nanoparticle solutions. We are grateful to Nicolas Faure, research engineer in LabHC, for technical assistance. We also thank the Centre Informatique National de l'Enseignement Supérieur (CINES, project A0180814604) for providing computational resources for the theoretical modeling aspects of this work.